\documentclass[12pt,latexsym]{amsart}
\usepackage{url}

\begin{document}
\title{Symplectic Quantum Mechanics and Chern-Simons Gauge Theory II:
Mapping Tori of Tori}

\author{Lisa C. Jeffrey}
\address{Department of Mathematics \\
University of Toronto \\ Toronto, Ontario \\ Canada}
\email{jeffrey@math.toronto.edu}
\urladdr{\url{http://www.math.toronto.edu/~jeffrey}} 

\begin{abstract}
We compute the semiclassical formulas for the 
partition functions obtained using two different Lagrangians:
the Chern-Simons functional and the symplectic 
action functional.
\end{abstract}

\maketitle
%\input mypream.tex

% Setting the display
%Commented out by LCJ to try to get to work
\def\baselinestretch{1.5}

\setlength{\parskip}{0.2\baselineskip}

\newcommand{\bC}{{\bf C}}
\newcommand{\bZ}{{\bf Z}}
\newcommand{\brpr}{\bar{\partial}}
\newcommand{\pr}{\partial}
\newcommand{\al}{\alpha}
\newcommand{\ga}{\gamma}
\newcommand{\Ga}{\Gamma}
\newcommand{\eps}{\epsilon}
\newcommand{\la}{\lambda}
\newcommand{\La}{\Lambda}
\newcommand{\om}{\omega}
\newcommand{\Om}{\Omega}
\newcommand{\Si}{\Sigma}
\newcommand{\si}{\sigma}
\newcommand{\nb}{\nabla}
\newcommand{\tlnab}{\tilde{\nabla}}

\newcommand{\lrar}{\longrightarrow}
\newcommand{\Tr}{\,{\rm Tr}\,} 
\newcommand{\End}{\,{\rm End}\,}
\newcommand{\Hom}{\,{\rm Hom}\,}
\newcommand{\Ker}{ \,{\rm Ker} \,}
 
\newcommand{\bla}{\phantom{bbbbb}} 
\newcommand{\onebl}{\phantom{a} }
\newcommand{\eqdef}{\;\: {\stackrel{ {\rm def} }{=} } \;\:}
\newcommand{\sign}{\: {\rm sign}\: }
\newcommand{\sgn}{ \:{\rm sgn}\:}                                      
\newcommand{\half}{ {\frac{1}{2} } }
\newcommand{\vol}{ \,{\rm vol}\, }

% define abbreviations for most common commands
%

\newcommand{\beq}{\begin{equation}}
\newcommand{\eeq}{\end{equation}}
\newcommand{\beqst}{\begin{equation*}}
\newcommand{\eeqst}{\end{equation*}}
\newcommand{\barr}{\begin{array}}
\newcommand{\earr}{\end{array}}
\newcommand{\beqar}{\begin{eqnarray}}
\newcommand{\eeqar}{\end{eqnarray}}
\newtheorem{theorem}{Theorem}[section]

\newtheorem{lemma}[theorem]{Lemma}
\newtheorem{prop}[theorem]{Proposition}
\newtheorem{definition}[theorem]{Definition}
\newtheorem{remit}[theorem]{Remark}
\newenvironment{rem}{\begin{remit}\rm}{\end{remit}}

\newcommand{\matr}[4]{\left \lbrack \begin{array}{cc} #1 & #2 \\
     #3 & #4 \end{array} \right \rbrack} 
\newcommand{\colvec}[2]{\left \lbrack \begin{array}{cc} #1  \\
     #2  \end{array} \right \rbrack}

\newcommand{\aff}{{\mbox{$\mathbb {A}$}}}
\newcommand{\RR}{{\mbox{$\mathbb {R}$}}}
\newcommand{\CC}{{\mbox{$\mathbb {C}$}}}
\newcommand{\ZZ}{{\mbox{$\mathbb {Z}$}}}
%\newcommand{\QQ}{{\mathbb {Q}}}
%\newcommand{\PP}{{\mathbb {P}}}

% calligraphic letters
\newcommand{\cala}{{\mbox{$\mathcal A$}}}
\newcommand{\calb}{{\mbox{$\mathcal B$}}}
\newcommand{\calc}{{\mbox{$\mathcal C$}}}
\newcommand{\cald}{{\mbox{$\mathcal D$}}}
\newcommand{\cale}{{\mbox{$\mathcal E$}}}
\newcommand{\calf}{{\mbox{$\mathcal F$}}}
\newcommand{\calg}{{\mbox{$\mathcal G$}}}
\newcommand{\calh}{{\mbox{$\mathcal H$}}}
\newcommand{\cali}{{\mbox{$\mathcal I$}}}
\newcommand{\calj}{{\mbox{$\mathcal J$}}}
\newcommand{\calk}{{\mbox{$\mathcal K$}}}
\newcommand{\call}{{\mbox{$\mathcal L$}}}
\newcommand{\calm}{{\mbox{$\mathcal M$}}}
\newcommand{\caln}{{\mbox{$\mathcal N$}}}
\newcommand{\calo}{{\mbox{$\mathcal O$}}}
\newcommand{\calp}{{\mbox{$\mathcal P$}}}
\newcommand{\calq}{{\mbox{$\mathcal Q$}}}
\newcommand{\calr}{{\mbox{$\mathcal R$}}}
\newcommand{\cals}{{\mbox{$\mathcal S$}}}
\newcommand{\calt}{{\mbox{$\mathcal T$}}}
\newcommand{\calu}{{\mbox{$\mathcal U$}}}
\newcommand{\calv}{{\mbox{$\mathcal V$}}}
\newcommand{\calw}{{\mbox{$\mathcal W$}}}
\newcommand{\calx}{{\mbox{$\mathcal X$}}}
\newcommand{\caly}{{\mbox{$\mathcal Y$}}}
\newcommand{\calz}{{\mbox{$\mathcal Z$}}}

\newcommand{\qqed}{\hfill \mbox{$\Box$}\medskip\newline}

\newcommand{\U}{U}
\newcommand{\tlbe}{{\tilde{\beta}}}
\newcommand{\ftil}{\tilde{f}}
\newcommand{\atil}{{\tilde{A}}}
\newcommand{\ad}{{\rm ad}}
\newcommand{\cs}{CS}
\newcommand{\lamax}{\Lambda^{\rm max}}
\newcommand{\lieg}{{\bf g}}
\newcommand{\liet}{{\bf t}}
\newcommand{\ddtau}{\frac{\partial}{\partial \tau}}
\newcommand{\nbt}{\nabla^{(t)} }
\newcommand{\bunt}{\widetilde{TN}}
\newcommand{\buntpr}{\widetilde{T^*N}}
\newcommand{\ddt}{\frac{\partial}{\partial t}}

\newcommand{\barpit}{{\bar{\Pi_t}}}

\newcommand{\tder}{\frac{d}{d t}}
\newcommand{\barpi}{{\bar{\Pi}}}
\newcommand{\tlu}{{\tilde{u} }}
\newcommand{\dbar}{\bar{\partial} }
\newcommand{\vardelta}{\delta}
\newcommand{\buntdopr}{\tilde{T''N }}

\newcommand{\tld}{\widetilde{D}}
\newcommand{\sigone}{\Sigma_1}
\newcommand{\sigthr}{\Sigma_3}
\newcommand{\sigtwo}{\Sigma_2}
\newcommand{\SF}{SF}
\renewcommand{\Tr}{{\rm Tr}}
\newcommand{\lineb}{{\mathcal{L}}}
\newcommand{\tf}{\tilde{f}}
\newcommand{\tlineb}{\tilde{\lineb}}
\newcommand{\hatx}{E}
\newcommand{\A}{A}
\newcommand{\abs}[1]{||#1||}

\newcommand{\rootl}{\Lambda^{\rm R}}
\newcommand{\weightl}{\Lambda^{\rm W}}

\newcommand{\N}{V}
\newcommand{\splin}{SL(2, \ZZ)}
\renewcommand{\frak}{\mathfrak}
\newcommand{\factr}{ (p-w-w^{-1}) }
\newcommand{\absc}{|c|}
\newcommand{\vpm}{|d + a \pm 2|} 
\newcommand{\posrts}{|\Delta_+|}

\newcommand{\htil}{{\tilde{H}}}
\newcommand{\hsig}{\calh(\Sigma)}
\newcommand{\lf}{{\call}_f}
\newcommand{\dph}{\dot{\gamma}}
\newcommand{\tilu}{{\tilde{u}}}
\newcommand{\dps}{\dot{\psi}}
\newcommand{\liesp}{{\mathfrak{sp}}}
\newcommand{\txomf}{T_{x_0} \Omega_f}
\newcommand{\txom}{{T_{x_0} \Omega^0}}
\newcommand{\ddu}{\frac{\partial}{\partial u}}
\newcommand{\tlxt}{\widetilde{X}_t}
\newcommand{\sympl}{Sp(2n, \RR)}
\newcommand{\rea}{\RR}
\newcommand{\zbar}{{\bar{z}}}
\newcommand{\zbari}{\zbar_i}
\newcommand{\zbarj}{\zbar_j}
\newcommand{\keradd} {{\rm Ker (ad)}}
\newcommand{\imadd}{{\rm Im(ad)}}
\newcommand{\coe}{\lambda}
\newcommand{\xhat}{E}
\newcommand{\zj}{Z_j}
\newcommand{\zk}{Z_k}
\newcommand{\jm}{j}
\newcommand{\tpr}{t'}
\newcommand{\dfn}{:=}
\newcommand{\bps}{\bar{\psi}}
\newcommand{\expad}{{\rm exp(ad) }}
\newcommand{\jb}{\bar{j}}
\newcommand{\sfd}

\newcommand{\be}{\beta}

\pagestyle{plain}

\tableofcontents
\section{Introduction} \label{intro}

\subsection{Chern-Simons gauge theory} \label{intro:csgt}
This article  is a companion to \cite{sqpaper1}. It treats  
 Chern-Simons gauge theory and a second theory,
symplectic quantum mechanics (SQM),
a field theory in $0+1$ dimensions for which the 
Lagrangian is the symplectic action functional. 
This paper treats the special case of 
mapping tori of tori.
We shall introduce a Lie group $G$ (usually $G=SU(2)$) and
an integer parameter $k$ (the level).
The background for this paper is as in \S1 of \cite{sqpaper1}.

\subsection{Summary of results}
Our work        treats the Chern-Simons partition function 
for mapping tori  $\Si_\be$ of surfaces $\Si$. 
In this article we  
treat only the case when $\Si$ is a 2-torus. The main
aim here is to use the reciprocity formula for Gauss sums to demonstrate
the large $k$ limit (\ref{csstph}) of the Chern-Simons partition 
function for mapping tori of tori. As a complement to 
\cite{sqpaper1},  we also compute 
the SQM partition function explicitly and show the terms in it 
agree with those in the rigorous large $k$ limit of the Chern-Simons 
partition function. In this 
case, a formal argument 
% similar to that in Proposition\  \ref{stexact}
shows the stationary phase approximation to 
the Chern-Simons partition function is exact 
(i.e.,  the action is precisely quadratic.) 
This is confirmed by our rigorous calculations in  \cite{lenspa},
recalled in \S \ref{gausder}.

Here
we  restrict the explicit treatment of symplectic
quantum mechanics to mapping tori of tori.
When $\Sigma$ is a torus, the stationary phase approximation 
to the symplectic quantum mechanics path integral is indeed
exact. We may thus expect exact 
agreement with the value for the partition function, 
rather than agreement only with the leading term in an 
asymptotic expansion.
We present two computations. For $G = SU(2)$, we have treated
an arbitrary element $\U$ of $SL(2, \ZZ)$. For arbitrary 
classical groups $G$, we have for simplicity treated 
only $\U = T^p S$  (see (\ref{sdef}) and (\ref{tdef}) below).

The notation in this paper is as in \cite{sqpaper1}.
The remainder of this article is organized as follows.
In \S  \ref{maptori} we specialize to mapping tori 
of tori. We prove (formally) the exactness of stationary
phase (Proposition\  \ref{pr3:stex}), and explicitly evaluate 
the quantities appearing in the stationary phase 
approximation of SQM. %There is also a discussion 

\S  \ref{s:sqmsu2} computes the SQM partition 
function. 
In \S  \ref{s:fram}, we discuss the natural framings
of mapping tori of tori, and how they enter in the 
comparison of the Chern-Simons and SQM partition functions.
\S \ref{gausder} recalls results from \cite{lenspa} which give the
large $k$ limit of the Chern-Simons partition 
function using Gauss sums.
In fact this calculation demonstrates that
the large $k$ expression gives the exact value, in 
accordance with our formal argument in \S  \ref{maptori}.

The main result in this article is
the
 comparison of the semiclassical formula for SQM with the 
semiclassical formula for Chern-Simons gauge theory of a mapping
torus of a torus (the critical points of both Lagrangians correspond
to fixed points of the diffeomorphism defining the mapping torus)
-- see Proposition   \ref{eq3:fp8} and (\ref{su2:trac}).

{\em Remark:} Much of  the material in this article
derives from the author's D. Phil. thesis \cite{J:thesis}.
Some 
other results from this thesis have already been published in \cite{lenspa}.

\section{Mapping tori of tori} \label{maptori}
\subsection{General method} \label{maptorgen}
The case of mapping tori of tori is  simpler in many respects than
the general case, as described 
in the introduction. Examples of three-manifolds 
$\Si_\be$ for $\Si$ a torus may be obtained by $0$-surgery 
on fibred knots whose fibre is a punctured torus: some of these
are
discussed in \cite{KK}. 

One difficulty with the torus case is that the moduli 
space $\calm(\Si)$ is singular, as is obvious from its 
description as a quotient space $(T \times T)/W$. Some of 
the fixed points of the symplectic diffeomorphism 
$f$ of $T \times T/W$ are in fact singular points, 
notably the product connection. It is not obvious how 
to rigorously define the quantities $\det D_x$ and 
$\Tr \: \ftil_x$ appearing in the contribution to 
the SQM partition function from such fixed points 
$x$. These points are points   fixed by some nontrivial 
$w \in W$: for $SU(2)$ they are the central flat connections. 

We shall circumvent this difficulty by working on $T \times T$. 
Observe that if $ A\in T \times T$ is a fixed point of a linear 
map $U$ acting on $T \times T/W$, then there is some 
$w \in W $ such that $A$ is a fixed point of $w U $ acting 
on $ T \times T$. The SQM data $\det D_A$ and 
$\Tr \: \ftil_A$ for the map $f = U$ 
on $T \times T/W$ thus naturally correspond to the SQM data
for $f = wU$ on $T \times T$. (Here, 
$D_A$ is the operator $ - \half J (\tder + \hatx)$ 
defined in \cite{sqpaper1}) It is thus natural to consider 
all such maps $f = wU$ on $ T \times T$ and sum their 
SQM partition functions. In fact, $A$ solves 
$w U A = A $ in $T \times T$ if and only 
if $ w'A$ solves $w'w (w')^{-1} U (w'A) = w'A$, so it is necessary
to divide our fixed point sum by $\vert W\vert$.

This prescription determines $ \Tr  \ftil_\A$ and $ \det D_\A$ 
for $\A \in T \times T$, provided $A$ is fixed by only one of 
the maps $w U$, i.e., provided no nontrivial  element 
of $W$ fixes $\A$. For those elements $\A$  which are fixed by some 
element in $W$, we sum the contributions to the SQM partition 
functions for all $wU$ that fix $A$. 

The regularization procedure for the eta invariant 
for a general moduli space (see \cite{sqpaper1}) 
enables us to replace the difference of eta invariants 
$\eta (A_+) - \eta(A_-)$ by a shift in the coefficient 
of the symplectic action functional $S(A_+) - S(A_-)$ 
from $k $ to $k+h$, plus a term involving the spectral 
flow mod $4$ of the path of operators between $D_{A_+}$ 
and $D_{A_-}$ associated to the gradient of the 
symplectic action functional. 
If $\A_+$ and $\A_-$ $ \in T \times T$ 
are fixed by $w_+ U$, $w_- U$ where $w_+$ and 
$w_-$ are {\it different}, we cannot naturally 
define the spectral flow between $D_{\A_+}$ and 
$D_{\A_-}$. We can, however, always define the 
spectral flow of the family of operators associated to the gradient of 
the symplectic action functional 
for $f = wU$ from a fixed point $A$ of $f$ 
to the product 
connection $A_0$, since $A_0$ is fixed by $wU$ 
for all $w$. In fact, it turns 
out (see Proposition\  \ref{sfsame}) that this 
spectral flow is zero. We thus just  need an ansatz to replace the 
spectral flow between the operator $D_{A_0} (wU)$ and 
the operator $D_{A_0} (U)$. This ansatz (based on 
the results of our rigorous  calculations using 
Gauss sums) is 
\beq \label{sfansatz}
\SF \left ( D_{A_0} (U), D_{A_0} (wU) \right ) 
 = 1 - \det w  \pmod{4}. \eeq

This gives the formula 
\beq \label{sqmz} 
Z_{SQM} (U,k) = i^\mu  \, \, \frac{1}{\abs{W} } \:  
\sum_{w \in W} \sum_{\stackrel{\A \in T \times T:}{wU \A = \A}}
\, \det w \, \frac{{\Tr \ftil_\A}^{k+h}  } 
{ {\abs{\det D_A(wU)} }^{1/2 } }.  \eeq
Here, $\mu(A_0)$ is the \lq\lq defect'' resulting from  a 
certain integer choice. %  (see (\ref{finregsqm}).)
This is (5.11) in \cite{lenspa}, 
which requires Conjecture 5.8 in that paper. This
conjecture has been proved by Himpel in 
\cite{Himpel}.

\begin{rem} \label{ambigsign} Even without the factor $i^\mu$,
 the overall sign of the SQM partition function
(\ref{sqmz}) is ambiguous: if $-1 \in W $, then under 
replacement of $U $ by $ - U $, the SQM partition function
changes by $\det w $, which is $1$ or $-1$.
\end{rem}

\subsubsection{Regularization of eta invariant: the torus case}
In regularizing the eta invariant in SQM for a symplectic 
manifold $N$, we obtained a correction term $-2 (\frac{i }{2 \pi}
\int_\tlu \Tr F_\nb )$, where  $ \Tr F_\nb$ was the curvature of the 
canonical bundle $\calk$ of $N$, viewed as a bundle  over the 
mapping torus $N_f$. We replaced this by $2 \int_u \al$, where 
$\al$ was a 2-form on $N$ representing the cohomology 
class $c_1(\calk)$.

In the case when $N = T \times T/W$, one must make sense of 
the canonical bundle of $T \times T/W$ rather than 
using the canonical bundle of $T$ (which is of course
trivial). Observe that for a branched covering 
$N \stackrel{\pi}{\to} M$ 
of complex manifolds, we have the \lq\lq Hurwitz formula'' 
\beq  \pi^\ast c_1 \calk_M + D = c_1 \calk_N , \eeq
where $D$ is the divisor corresponding to the 
branch locus (taken with some multiplicity). 
For us, $N = T \times T$, and $M = T \times T/W$ 
is no longer a manifold. Nonetheless, we adopt this as 
a definition of $\pi^\ast c_1 (\calk_M)$; this yields 
(see \cite{ADW}, (5.30))
$$\pi^\ast c_1 \calk_{T \times T/W} = 2h \Bigl (\frac{ \om}{2 \pi}
\Bigr ), $$
where $\om$ is the basic symplectic form. This leads as before 
to the shift of the coefficient of the symplectic action functional 
from $k$ to $k + h$.

\subsection{Lifting $f$ to the prequantum line bundle}

We now discuss the SQM data for $\calm$ when $\Si$ is a 
torus.
% we refer to \S  \ref{ss:thqt}, 
We may view $\liet \oplus \liet $ as a subspace of the space of
connections $\cala$ on $\Si$, and the actions of 
$W$ and $\La = \rootl \oplus \rootl$ as gauge 
transformations. It is easy to 
check by explicit calculation that our lifting of 
these actions to $\call$ %that was described in \S \ref{ss:thqt}
coincides with the lifting via the Chern-Simons functional described
in \cite{sqpaper1} (see (68) in that paper).

\noindent{\bf Notation:} The diffeomorphism $\be$ of $\Si$ corresponds
to an element $U \in SL(2, \ZZ). $ We shall write $f$ 
or $f_U$ for the corresponding map on 
$T \times T/W$, and $\ftil$ or $\ftil_U$ for 
its lift to the prequantum line bundle $\call$ over 
$T \times T/W $ or $T \times T$.

We now choose a lift of $f_U: \calm \to \calm$
to $\call$, preserving the connection. We choose the trivial 
lift to the trivial bundle over $\aff$:
\beq \label{eq2:ftil}
\tilde{f}_U(A, z) = (UA, z) \eeq
This is easily shown to preserve the connection
on $\call$. 
However, all lifts to the prequantum line bundle preserving 
the connection coincide up to a constant in $U(1)$. 
That the lift (\ref{eq2:ftil})
coincides {\it precisely} with the lift 
using the Chern-Simons functional follows from 
the fact that they agree on the product connection $A = 0$.

We need to choose a lift of $\be: \Si \to \Si$ 
to $\tlbe: P_\Si \to P_\Si$, and a flat  connection 
$A_0$ preserved by $\tlbe$. We do this by choosing
a trivialization of $P_\Si$ and letting
$A_0$ be the product connection and $\tlbe$ the 
trivial lift. This choice of $\tlbe$ then preserves 
the subspace $\aff$ of connections with constant 
coefficients in $\liet$. We identify $A_0$ with
$0 \in \liet;$ this enables us to lift 
the action of $\tlbe$ on $T \times T$ to the
linear action of $U$ on $\aff$. Of course the connection 
on the symplectic affine space $\aff$ 
is simply the restriction 
to $\aff$ of the connection
defined in \cite{sqpaper1}.
%%(\ref{3.2:conncal}).

We now show 
\begin{lemma} \label{equivar}
The lift of $U \in SL(2,\ZZ) $
given by  (\ref{eq2:ftil}) 
and our lift of $ w \in W$ 
%given in  
%\S  \ref{ss:thqt}  
are equivariant with respect to 
 the action of the lattice $\La$ 
on $\call$.
% which was defined in 
%(\ref{ch2:laact}).
\end{lemma}
\noindent{\bf Proof:} Write $\N$ for the corresponding linear maps
on $\liet \oplus \liet.$ 
The equivariance   condition is characterized by the following 
equation on $\call = \aff \times \CC$: 
$$(\N A + \N \lambda, \,e_\lambda (A) v) 
= (\N A + \N \lambda, e_{\N (\lambda) }(\N A) \,v). $$
Now 
$$\frac {e_{\N (\lambda)} (\N A)   }{e_\lambda(A) } = 
\frac{\epsilon(\N \lambda)}{\epsilon(\lambda)}, $$
so we need $\epsilon(\N \lambda )  = \epsilon(\lambda)$. 
Actually we need only check this for $\lambda$ in some basis of 
lattice vectors. 

We fix the coroot  basis  $ \{ h_\al \}$    of $\rootl$, and correspondingly
a basis     $ \{ h_\al^{(1)}, h_\al^{(2)} \} $ of 
$\rootl \oplus \rootl $. We define the theta-characteristic 
by 
\beq \label{thchardef}
\epsilon (h_\al^{(i)} ) = 1. \eeq
Then for $U \in \splin$, $U {h_\al}^{(i)} = 
m h_\al^{(1)} + n h_\al^{(2)} $ for some $m, n \in \ZZ$; because 
$h_\al$ does not mix with the other coroots 
$h_\beta, \beta \ne \al $ under $U$   and because $\langle h_\al, 
h_\al \rangle \in 2 \ZZ$, we have $\eps(U h_\al^{(i)} )  = 1$. 

Similarly for 
$ w \in W$, $w \,h_\al^{(i)}  = \sum_\beta  n_\beta h_\beta^{(i)}$: the 
two summands 
$\liet_1 $ and 
$\liet_2$ in $\liet \oplus \liet $ do not mix. So since $\omega$ 
pairs $\liet_1$ with $\liet_2$, again $\epsilon (w \, h_\al^{(i)} ) 
= 1. $ Hence the definition 
(\ref{eq2:ftil})              does 
indeed give a lift to $\call$.  \qqed

\begin{rem} {\bf : Theta-characteristics}

The choice of a theta characteristic for a 
bundle $\call$ on $T \times T$ is the specification 
of $w_1 (\call)$ $\in H^1(T \times T, \ZZ_2)$. 
We know that $ (T \times T)/W$ is simply connected, 
so bundles with different choices of theta-characteristic
on $ T \times T$ descend to isomorphic bundles 
on $ T \times T/W$, and the choice of theta-characteristic 
is irrelevant for our purposes. For convenience in specifying
the lift of $SL(2, \ZZ)$ to $\call$, we make the particular 
choice (\ref{thchardef}) for the theta characteristic. 
A different choice would force us to  choose a different lift 
in order to make it equivariant with respect to the $\La$ action.

The theta characteristic we have chosen is obviously identically $1$
in the $SU(2)$ case. \qqed   \end{rem}

\subsubsection{Stationary phase approximation }
\begin{prop} \label{pr3:stex}  
The stationary phase approximation for the SQM partition 
function corresponding to the moduli space of flat connections
on a torus is exact.
\end{prop}

\noindent{\bf \lq\lq Proof'' (formal):}
We view $T \times T$ as 
$ (\liet \oplus \liet) /( \rootl \oplus \rootl) $.
A basis of $\liet$ then defines coordinates on $T \times T$, 
in which
the symplectic
form $\omega$ on $\liet \oplus \liet$ is  a 2-form 
with constant coefficients and
the diffeomorphism $f$ is a linear map. The Lagrangian is defined
by parallel transport in $\call$ around a path $\ga$ with 
$f (\ga(0))  = \ga(1).$ Near a critical point $x_0$ this is 
given by 
$$L(\ga) = \int_u \, \omega $$
where $u(t,\tau): \RR \times I \to N$ is a homotopy from 
the constant path  $\ga_{x_0} $ to $\ga$ with $ u (t+1, \tau) = 
f ( \,u(t,\tau)\,)$. Because $f$ is linear, we may take 
$u(t, \tau) = \tau \ga(t)$. Because $\omega$ has constant
coefficients,
the integral becomes 
$$\int_{I \times I} \tau\omega(\ga, \dot{\ga}) \, dt \, d\tau= 
\frac{1}{2} \int_I \omega (\ga, \dot{\ga} ) \, dt, $$
which is precisely quadratic: hence the stationary phase 
approximation is 
exact.\qqed

\subsection{Fixed points of {\it f}, and action at the fixed points}
\label{ss:fixact}
 
If $A$ is a fixed point of $f$ on $T \times T/W$, there are
$ w \in W$ and $\lambda \in \Lambda$ such that 
\beq w U A - A = \lambda 
\eeq
in $\liet \oplus \liet.$ 
  The trace 
of $\ftil$ at a fixed point is computed as follows:
\begin{eqnarray*} 
\widetilde{w f} (A,v) 
=& (wU (A),  v ) = (A + \lambda,  v)   \\
                   =& (A, e_\la(A)^{-1}  v ) ,
\end{eqnarray*}
%the last step using (\ref{ch2:laact}).
In other words 
\begin{equation} \label{eq2:tracef}
{\rm Trace} \, \tilde{f}\vert_A = e_\lambda(A)^{-1}     \end{equation}
$$
     \bla =  \exp \frac{i}{2} k \omega(A, \lambda) \epsilon^k(\lambda)  . \\
\bla 
$$
Explicitly, 
\begin{lemma}
The fixed points $A_\la$ are in correspondence
with $\la = (\la_1, \la_2) \in \La/(w U - 1) \La$: we define 
$$A_\la = \colvec{A_1}{A_2} = (wU-1)^{-1} \la.  $$ Furthermore,
the trace of the lift $\ftil$ at the fixed point $A_\la$
is given by   
\beq \label{3.3:tracetor}
{\rm Trace} \ftil\vert_{ A_\la  } = \exp \frac{i k  }{2} \: 
\om \left (  
[wU - 1]^{-1} \la, \la \right )  \eps^k(\la), \eeq 
$$ = \exp \left ( - i k \pi \left \langle  [wU-1]^{-1} \la,
S \la \right \rangle  \right )    \eps^k(\la)\bla  $$

where the theta-characteristic $\eps(\la)$ is defined by 
(\ref{thchardef}).  $\square $
\end{lemma}

By the  discussion after (\ref{eq2:ftil}), we actually have that 
\beq {\rm Trace} \ftil \vert_A = e^{2 \pi i \cs(\atil) }, \eeq
where $\atil$ is the flat connection on $\Si_\be$ corresponding
to $A$. This proves:

\begin{theorem} 
The Chern-Simons invariant of the flat connection 
$A_\la$ on the mapping torus $ \Si_U$ of the torus $\Si$ 
is:
$$\cs (A_\la) = \frac{1}{4 \pi}\: \om \left ( \, (wU - 1)^{-1} 
\:\la\; , 
\:\la\, \right ) + 
\begin{cases} 0, &\text{if  $ \eps(\la ) = 1;$} \\ 
			\frac{1}{2}, &\text{if $ \eps(\la ) = -1.$}   
\end{cases}
\square  
$$
\end{theorem}
\noindent Kirk and Klassen (\cite{KK}, Th. 5.6) have obtained this 
result for $G = SU(2)$.

\subsection{Absolute value of determinant}
As discussed in \cite{sqpaper1}, 
for the SQM operator $D$, the
value of $\vert \det D \vert^{-1/2} $ is $ \vert \det (df - 1)
\vert^{-1/2}, $ where $df: T_x M \to T_x M$ at the fixed point $x$
of $f$. 
For $ M = T \times T/W$ this becomes 
\beq \label{3.3:detval}
\vert \det D \vert^{-1/2} = \vert \det (wU - 1) \vert^{-1/2},  \eeq
for $w, U$ acting on $ \aff $.   

\subsection{Spectral flow}

Recall that $D$ was $ - \frac{J}{2} (d/dt + \hatx)$, where 
$\hatx$ was chosen in $\frak{sp}(2n) \otimes \CC$ such 
that $\exp \hatx = w f. $ %
 Above (\ref{sfansatz}), we have discussed the ansatz to make sense 
of the spectral flow between  fixed points of $U$ and 
fixed points of $w U$, when $w \ne 1.$ Here we show the following:
\begin{prop} \label{sfsame}
Consider the spectral flow of the operator $\cald$ 
corresponding to the gradient of the symplectic 
action functional. Consider a fixed value of $w \in W.$ 
 Between two fixed points $x_+, x_-$ 
of $ w U$ in $T \times T$, this spectral flow  is 0. 
\end{prop}

\noindent{\bf Proof:} The spectral flow between $x_+$ and $x_-$ 
is the difference of {\it Maslov indices} $\mu(x_+) - 
\mu(x_-) $. The Maslov index $\mu(\Psi)$ is 
associated to a path $\Psi$ in $Sp(2n, \RR) $.
Appropriate paths $\Psi_\pm$ are obtained from a trivialization of 
the tangent bundle (satisfying 
appropriate periodicity conditions) 
over a strip joining the two fixed 
points $x_+$ and $x_-$. 
(For details see \cite{DS} after equation (4.4).) 
Using the canonical trivialization of the tangent 
bundle of $ T \times T$ and a path of linear symplectic 
maps $f_t \in SL(2,\RR) $ joining $1$ and $U$, one   
easily sees that a trivialization may be constructed 
so that the paths $\Psi_\pm$ are the same. Thus 
the associated  Maslov indices are the same. \qqed

\section{SQM partition function } \label{s:sqmsu2}

\subsection{{\it G = SU(2)}}
We shall present the calculation of the SQM partition function 
for $T \times T/W$. We denote by $U$ an arbitrary element 
$$U = \matr{a}{b}{c}{d} $$
of
$SL(2, \ZZ)$ (provided $U$ is not parabolic, i.e., $\Tr (U) \ne \pm 2$).

We wish to compute the SQM formula (\ref{sqmz})
for the mapping torus partition function.
From  (\ref{3.3:detval}), we need the quantity 
$$\det (wU - 1) = 2 \mp (a + d). $$
Equation
  (\ref{3.3:tracetor}) allows us to 
evaluate the action:

$$(wU - 1)^{-1} = - \frac{1}{a + d \mp 2} \matr{d \mp 1 } {-b}{-c}{a \mp 1},
$$ 

$$ L(A_\la ) = -  \pi \left \langle (wU - 1)^{-1} \la, S \la \right 
\rangle $$
$$ = -  \pi \left \langle \la,( w U^t - 1)^{-1} S \la \right \rangle, $$
where 
\beq \label{sdef} S = \matr{0}{-1}{1}{0} \bla \in SL(2, \ZZ)\eeq
and 
\beq \label{tdef} T = \matr{1}{1}{0}{1} \bla \in SL(2, \ZZ)\eeq
descend to generators of the group $PSL(2,\ZZ)$.
(Here, $\langle \cdot , \cdot \rangle $
is the inner product on 
$\liet \oplus \liet$, and 
$S \in SL(2, \ZZ) $ acts on $\liet \oplus \liet$.)
For $SU(2) $ this becomes  
 \beq L(A_\la) = \frac{2 \pi }{a + d \mp 2} 
\left ( - c \la_1^2 + b \la_2^2 
+ (a-d) \la_1 \la_2\right ), \eeq
where we have used the coroot basis and the inner product
to identify $\rootl$ with 
$\ZZ$.
Our expression then reads 
\begin{eqnarray} \label{su2:sqm}
Z &=&  \frac{1}{2} \,\,i^\mu \,\, \sum_{\pm} \,\sum_{\la} 
\pm \: \frac{1}{\sqrt{\vert a + d \mp 2 \vert } } \: \times \hfill
\onebl   \\
\onebl &\onebl& \hfill \times \exp \left \{
2 \pi i \frac{(k+h) }
{a + d \mp 2}  \left ( - c {\la_1}^2 + b {\la_2}^2 + (a-d) \la_1 \la_2
\right ) \right \},  \nonumber \end{eqnarray}
where we sum over $\la = (\la_1,\la_2) \in \La/(\pm U - 1) \La.$
\subsection{ General {\it G} }

In the case of general $G$ we shall restrict ourselves 
for simplicity to one specific family, namely 
those $U \in SL(2, \ZZ)$ for which 
$c = 1$.
 
\noindent{\bf Notation:} $p$ will denote $\Tr (U)$. 
 
\begin{lemma} \label{prevlem} We have
$$ \vert \det(w \otimes U - 1) \vert = 
\vert \det( \Tr( U) - w - w^{-1}) \vert. $$
\end{lemma}

\noindent{\bf Proof:} If the eigenvalues of $U$ are $\la, \la^{-1}$ 
and those of $w$ are $\mu$ then this breaks up as
$$ {\rm LHS } = \left \vert \prod_\mu (\la \mu - 1) \,
(\la^{-1} \mu - 1)\, (\la \mu^{-1} - 1) \,
(\la^{-1}\mu^{-1} -1)\, \right \vert^{1/2} $$
$$=  \left \vert \prod_\mu  (\la^{1/2}\mu^{1/2} - \la^{-1/2}\mu^{-1/2} )\,
(\la^{-1/2}\mu^{1/2} - \la^{1/2}\mu^{-1/2} ) \right \vert $$
$$ =  \left \vert \prod_\mu (\la + \la^{-1} - \mu - \mu^{-1}) \:
\right \vert  = {\rm RHS}. \bla \square$$

\begin{lemma}
A basis of representatives for $\La/(wU - 1) \La $  is 
given by $$(\si,0),  \bla \si \in \rootl/(p - w - w^{-1} )\rootl. $$
\end{lemma}

\noindent{\bf Proof:} 
By Lemma \ref{prevlem}, these sets have the same 
number of elements. Now 
$$(wU-1) \La = (U - w^{-1}) \La, \bla \mbox{where} $$
$$U - w^{-1} = \matr{a - w^{-1}} {b}{c}{d - w^{-1} }. $$
As $c = 1$, there is clearly a basis of representatives 
of the form $(\si, 0)$ (since there is an element of 
$ (U-w^{-1}) \La$ of the form $(n,1)$).  Also, 
$$\matr{a - w^{-1} } {b}{c}{d - w^{-1} } \colvec{ - (d - w^{-1} )\si }
{\si}  = \colvec{ [p - w^{-1} - w ] w^{-1} \si} {0}, \bla $$
so for any $\si \in \La^R$, $[\:(p-w-w^{-1} )\si, 0 ] \in 
(wU-1) \La.$ 
$\square $ 

\noindent{\bf Remark:} As a set of 
representatives $\la \in \La$ can be chosen in 
this way, with the second 
component $\la_2$ 
equal to zero, it is easy to see that the theta characteristics
$\eps (\la)$ can be chosen as 1.  

We need the factor 
$$L(A)  = - 2 \pi \langle  A_\la, S \la \rangle =  2 \pi 
\langle A_2, \si \rangle, $$
where $(w U - 1) A = \la$. 
Explicitly, 
$$\matr{(aw-1)}{bw}{cw}{(dw-1)}  \colvec{A_1}{A_2} = 
\colvec{\si}{0}, $$
so substituting for $A_1$, we obtain 
$$A_2 = (w + w^{-1} - d - a)^{-1} \si  .  $$

Thus we have 
\begin{prop} \label{eq3:fp8}
The overall SQM formula for the mapping torus
partition function is
\begin{eqnarray*}
Z_{SQM}(U,k)  &=&  
i^\mu \, \frac{1}{\vert W \vert} \sum_{w \in W} \frac{ \det(w) }
{\sqrt{ \vert \det \factr\vert } } \: \times \hfill \onebl \\
\onebl & \onebl & \hfill \times \sum_\sigma \exp  \Biggl  \{
- r \pi i \Bigl \langle \,\factr^{-1} \sigma\,,\, \sigma 
\,\Bigr \rangle \: \Biggr \}.  \bla \square \end{eqnarray*}
\end{prop}
Here the second sum is over $\si \in \rootl/\factr \rootl,$
and $\mu$ is the \lq\lq defect'' from
\cite{sqpaper1} (see (68) in that paper).

\section{Framings of mapping tori}\label{s:fram}

The Chern-Simons-Witten invariant  of a 3-manifold $Y$
depended on 
the specification of a {\it 2-framing } (a trivialization
of $2TY$). For a discussion of how this dependence
shows up in the path integral, see \S 5.4 in  \cite{sqpaper1}
For mapping tori of surfaces $\Si$,  the 
possible 
  2-framings correspond   to 
maps from the mapping class group $\Ga$  
 to its $\ZZ$-central extension 
$\hat{\Gamma}$  
(see
\cite{A2}). One such  map $s$ corresponds 
to the canonical framing (see \S 5.4 in \cite{sqpaper1}) ).
If $\Si$ is a 
torus, 
the extension actually splits, so 
another framing is defined by the unique homomorphism   $s_1: \Ga \to 
\hat{\Ga}$.
The discrepancy between the framings $s$ and $s_1$ is 
discussed in \cite{A1}, \cite{A2}: it is identified as the 
signature defect of $\Sigma_U$ in the framing $s_1$ (as the framing $s$
is the one giving zero signature defect.) We have 
(\cite{A1}, 6.15
and 5.4):
\beq \label{diffram}
s_1(U) - s(U) = \Phi(U) - 3 \sign c(a+d)  , \eeq  %%%sign?
where 
and $\Phi(U)$ is the Rademacher phi function \cite{Rad}.
(Actually,
if $U$ is hyperbolic, the signature defect of the mapping 
torus $\Si_U$ is equal to its eta invariant in a natural 
metric: see \cite{A2}. In other words, in this case there
is a canonical metric for 
which the \lq\lq counterterm'' vanishes.)

If $U$ is hyperbolic, and conjugate to diag($e^h, e^{-h}$), then
(see \cite{A1}, (5.51)) the
framing $s_1$ on $\Si_U$ can be constructed quite 
explicitly: it corresponds to using 
diag($e^{th}, e^{-th}$) ($0 \le t \le 1$) to define a path 
in the space of framings on $\Si$, and hence a framing on 
$\Si_U$.    For simplicity, we restrict our comparison 
of phase factors to the case when $U$ is hyperbolic
(i.e., when $\abs{\Tr (U)} > 2 $.)

\begin{rem}  \label{frameass} {\it  Framing assumption:}
Note that the symplectic quantum mechanics partition 
function contained an overall fourth root of unity  $i^\mu$ 
corresponding to the choice of a trivialization. 
{\it The formulas  in this section are for the 
case when $i^\mu = 1.$
} \end{rem}

\subsection{{\it SU(2)} case}
 
Apart from the factor $i^\mu$ in (\ref{su2:sqm}),
the discrepancy between (\ref{su2:sqm}) and 
(\ref{su2:trac}) is a factor $$iK \zeta^{\sign c(d + a \pm 2)} $$ 
multiplying the trace partition function. If $U$ is hyperbolic, 
this is just $i \zeta^{\sign c(d+a)- \Phi} \sign c$.

Our discrepancy is

$$\zeta^{2 - \psi - 2 \sign c (d+a) + 2 \sign c - 2} 
\; = \; \sign(d+a)\zeta^{- \psi},$$
so we have

\beq \label{su2disc}
\Tr \calr(U) = \zeta^{-\psi(U)} \sign(d+a) Z_{SQM}(\Si_U). 
\eeq

\subsection{General {\it G} }

The phase discrepancy between 
the SQM result (Proposition\  \ref{eq3:fp8}) and the result
(\ref{eq4:trac}) for $\Tr \calr(U)$  is now 
$$ i^{\posrts} \: \exp \Bigl \{  - p i \pi \frac{\vert \rho \vert^2 }{h} 
\Bigr \} 
\;  
\exp i \pi l \frac{\sign \det (p - w - w^{-1})}{4} .$$
In the case when $\vert p \vert > 2$, this is

\beq \label{gen:calc}
 i^{\posrts} \: \exp \Bigl \{ - p i \pi \frac{\vert \rho \vert^2 }{h}
\Bigr \}  \; 
\exp i \pi l \frac{\sign p }{4} . \eeq
Using $\posrts = (\dim G - l)/2 $, 
this becomes 
$$ i^{\posrts} \: \exp \Bigl \{ -  2 \pi ip \frac{\dim G }{24} \Bigr \}\: 
\exp i \pi  \frac{l\sign p+ \dim G - l }{4} .$$
In this case, the expected  correction  factor 
caused by the framing 
is 
\beq \label{gen:exp}
\exp  \Bigl \{- \frac{2 \pi i  \psi(U) \dim G}{24} \Bigr \}, \eeq 
and $\psi(U) $ in this case is $p - 3 \sign p$.  
A short calculation shows that
equations (\ref{gen:exp}) and (\ref{gen:calc}) differ 
only by a sign $(\sign p)^{\posrts}.$ 

Thus, up to a sign, the difference between the trace
calculation and the SQM calculation for $i^\mu = 1$ 
is accounted for
by a change in framing, embodied in the 
factor $\psi$. 
The sign   ambiguity is to be expected from 
the definition of the SQM partition 
function (see Remark \ref{ambigsign}),
although we do not know how to resolve it.

\noindent{\bf Remark:} Notice that once the phase has been 
corrected (by the procedure for regularizing the 
Chern-Simons theory eta invariant, described in \S 5.4 of 
\cite{sqpaper1}, 
one obtains {\it exact}
agreement between the SQM result  
(or equivalently the stationary
phase expansion of Chern-Simons: see \S 4 of \cite{sqpaper1})
and
the result for $\Tr \calr(U)$. This is in contrast
to the lens space case (see 
\cite{lenspa}), where    one only obtains
asymptotic agreement with the stationary 
phase formula.

\section{Gauss sum derivation of Chern-Simons partition function} \label{gausder}
\subsection{{\it G = SU(2)}}

Equation   (4.7) of \cite{lenspa} reads as follows:

\begin{equation} \label{su2:trac}
\phantom{iii}
\end{equation}
\begin{eqnarray*} 
 Z(U,r) = \sum_\pm  &\pm& \,\frac{1} 
{2 i \absc\sqrt{\vert d + a \pm 2 \vert } } \:
K(U) \: \zeta^{\sign(c(d + a \pm 2) )} \:  \times \hfill  \bla  \\
\onebl \onebl   \hfill &\times&  \sum_{\beta \pmod{c} } 
\sum_{ \ga = 1}^{\vert d + a \pm 2 \vert}
\exp 2 \pi i r 
\frac{- c \gamma^2 + (a-d) \gamma \beta + b \beta^2 } { d + a \pm 2} . 
\nonumber \end{eqnarray*}
Up to the phase which was the subject of  \S   \ref{s:fram},
this is the SQM result (\ref{su2:sqm}). Note that the SQM  result was 
expressed as a sum over a fundamental domain of 
$\La$ under the action of $B = 1 \pm U$.  The equivalence
of this with (\ref{su2:sqm}) is established by the
following observations:
\begin{enumerate}
\item  $\det B  =  2 \pm a \pm d. $

 \item The sum 
\beq \label{sumcexp}
\sum_{\beta = 1}^{\absc \vpm} \sum_{\gamma = 1}^{\vpm}
\exp 2 \pi i r 
\frac{- c \gamma^2 + (a-d) \gamma \beta + b \beta^2 } { d + a \pm 2}  \eeq
equals $ \abs{\det B} $ times the sum in (\ref{su2:trac}).  
If $(a,b)$ of $\La \cong \ZZ^2$ is such that
$\det B$ divides $a$ and $b$, then $(a,b)$  is in $B \La$, so the 
points $(\beta, \gamma) = (0, \vpm)$, $(\beta, \gamma) = (\absc 
\vpm, 0)$ are in $B \La$. Hence (\ref{sumcexp}) 
covers precisely $\vert \det B \vert \absc$ fundamental domains
(as it covers an integer number of domains, each of which 
contains $\vert \det B \vert $ points). \qqed
\end{enumerate}
\subsection{General {\it G} }

The conjugacy classes of $U$ for which $c = 1$ can be 
represented by $U = T^p S $, where 
$T$ and $S$  are elements of $SL(2,\ZZ)$ 
which descend to  the standard generators of 
$PSL(2,\ZZ)$ (see (\ref{sdef}) and (\ref{tdef})).
We obtain

\beq \label{eq4:trac}
Z = {\rm Tr}(T^p S )  = 
 i^{\posrts} 
\left \vert \frac{{\rm vol} ( \weightl)}{r {\rm vol} \rootl } 
\right \vert^{1/2}  \exp \left \{ - \frac{p i \pi \langle \rho , 
\rho \rangle }{h}  \right \} \times
\eeq
$$ \times   \sum_{w \in W} \det(w) \sum_\lambda \exp \left \{
\frac{i \pi}{r} \left \langle \:\factr\: (\la + \rho)\: , \:  \la + \rho 
\:
\right \rangle \,\right \} ,   $$
where the sum is over $\la \in \weightl$ satisfying 
an integrality condition.

Let us analyse the symmetries of the trace sum (\ref{eq4:trac})
with a view to 
expressing it as a sum over $\weightl / r \weightl$. Define 
(for $\la \in \weightl$)
$$g(\la) = \sum_{w \in W} \det(w) \cdot  \exp \frac{\pi i\langle (p-2w) \la, 
\la \rangle }{r}. $$
The trace is obtained by summing $g(\la)$ over weights 
$$\{\la  = \mu + \rho\,: \:\mu \in \overline{{\rm FWC}}, 
\, \:\langle \mu, \al_m \rangle \le k\}, $$
or $$ Z = \sum_\la g(\la)$$   
         where the sum is over $\{\la \in {\rm FWC} \mid 
 \langle \la, \al_m \rangle  < k + h.\}$

The following result is \cite{lenspa}, Proposition 4.4:
\begin{prop} \label{3.6:invce}
$g(\la)$ is invariant under:
\end{prop}

(i) $\la \to  - \la$ (obvious)

(ii) $\la \to u \la, u   \in W:$ for
$$\left \langle \,u\la\,,\, (p-2 w) u \la \,\right \rangle  
= \left \langle \la,\, 
\{\,p - 2 (u^{-1} w u)\,\}  \,\la   \,
\right \rangle. $$

(iii)   $\la \to \la + r h_\al$, $\al$ any root ($h_\al $ denotes the 
corresponding coroot $2 \al /\langle \al, \al \rangle $).
For 
\begin{eqnarray*} 
\phantom{a}&\onebl & \bla   \frac{1}{r}  \Bigl \langle \,\la + r h_\al,\, 
(p-2w)\, \left (\,\la+  r h_\al\,\right ) \,\Bigr \rangle       \\ 
 = &\onebl &  \frac{1}{r} \langle \la \, , \,(p-2w)\, \la \rangle
+2 \langle\, h_\al\, , \,  (p-2w) \la \,\rangle
 +  r \langle \, h_\al \, , \, (p-2w) h_\al \, \rangle. 
\end{eqnarray*}
The second term is obviously in $2 \ZZ$, since 
$h_\al$ is in the integer lattice.  The third
term is also in $2 \ZZ$, since 
$\langle h_\al, h_\al \rangle  \in 2 \ZZ$ (a property of the basic
inner product), and 
$$\langle h_\al, w h_\al \rangle = \frac{2 \langle w \al, h_\al \rangle   
}{\langle \al, \al \rangle}. $$
%The highest root, a long root, is normalized to have 
%${\rm length}^2 = 2  $, and the length squared of short roots is then 
%$2/n$ for some $n \in \ZZ$. 

(iv) $g(\la) = 0$ for a weight $\la$ with $\langle \la, \al \rangle 
= r \,n$ for any root $\al$ ($n \in \ZZ$).

The following result is Proposition 4.5 in \cite{lenspa}:
\beq \label{eq6:trsum2} 
\Tr \, \calr (U) = 
\exp \left \{- \frac{p \pi i \langle \rho, \rho \rangle }{h}
\right \} 
\, i^{\posrts}  \, \exp \left \{ \frac{i \pi l \:{\rm sgn} \det(B) }{4}
\right \} 
\; \frac{ \vert \det(B) \vert^{-1/2}}{\vert W \vert} \times  \eeq
$$\times   \sum_{w \in W} \det(w) \sum_{\mu \in \rootl/B \rootl} 
\exp -i \pi \langle \mu, r B^{-1} \mu \rangle. 
$$
The last expression equals what we obtained 
(Proposition\  \ref{eq3:fp8}) from 
the fixed point calculation, up to the phase which we investigated in the
previous section.

\newcommand{\auth}{\sc}

\end{document}